\documentclass[twocolumn, showpacs, prl]{revtex4}

\usepackage{graphicx}

\usepackage{dcolumn}

\usepackage{bm}

\begin{document}

\title{Zeeman splitting of interacting two-dimensional electrons with two effective masses}

\date{\today}

\author{K.\ Vakili}

\author{E.\ Tutuc}

\author{M.\ Shayegan}

\affiliation{Department of Electrical Engineering, Princeton
University, Princeton, NJ 08544}

\begin{abstract}

We have realized an AlAs two-dimensional electron system in which
electrons occupy conduction-band valleys with different Fermi
contours and effective masses. In the quantum Hall regime, we
observe both resistivity spikes and persistent gaps at crossings
between the Landau levels originating from these two valleys. From
the positions of the spikes in tilted magnetic field and
measurements of the energy gaps away from the crossings, we find
that, after occupation of the minority valley, the spin
susceptibility drops rapidly, and the electrons possess a {\it
single} interaction-enhanced g-factor, despite the dissimilarity
of the two occupied valleys.

\end{abstract}

\pacs{73.21.Fg, 73.43.-f, 73.43.Qt, 73.61.Ey}

\maketitle

Coincidences between Landau levels (LLs) in multi-component
quantum Hall systems provide information regarding the relative
energy scales of the two-dimensional electron system (2DES). For
example, coincidences between LLs with different spin occur when
the ratio of Zeeman and cyclotron energies is an integer, yielding
the product of the interaction renormalized effective mass ({\it
m}$^*$) and Land\`{e} g-factor ({\it g}$^*$) which is proportional
to the spin susceptibility of the system \cite{fang68}. To date,
most studies of LL crossings have been conducted in systems
wherein the discrete components (e.g. two layers in a symmetric
bilayer) are very similar \cite{muraki01}. Here, we report the
realization of a 2DES in which electrons occupy two very
dissimilar valleys, exhibiting different Fermi contours and
effective masses. We study inter-valley LL crossings in this
system, manifested by resistivity spikes, and extract information
about the relative energy scales of the two valleys. We find that
{\it g}$^*${\it m}$^*$ drops rapidly upon occupation of the
minority valley and, remarkably, that the electrons possess a
single {\it g}$^*$ despite the dissimilarity of the two occupied
valleys.

Aluminum Arsenide (AlAs) is a practical system to study crossings
between LLs originating from different valleys thanks to the
degree of control that exists over valley occupation in this
material. Bulk AlAs has three conduction band minima at the
X-points of the Brillouin zone, each leading to an ellipsoidal
Fermi surface with a longitudinal mass of {\it m}$_{l}$ = 1.04 and
transverse masses of {\it m}$_{t}$ = 0.21 \cite{momose99}, in
units of the vacuum electron mass. Though confinement along the
[001] axis lowers the energy of the valley oriented out of the
plane (Z) with respect to the two in-plane valleys (X and Y), the
strain caused by lattice mismatch between the AlAs quantum well
(QW) and the GaAs substrate on which it is grown has the opposite
effect. Consequently, there is a crossover between Z and (X,Y)
valley occupation at a threshold QW thickness of $\simeq$ 55 $\AA$
\cite{vankesteren89, vandestadt96}. By studying a QW near this
thickness, we sought to simultaneously occupy the Z valley and
either the X or Y valley \cite{residual}. The former has an
isotropic Fermi contour projection in the plane of the 2DES with a
band effective cyclotron mass of {\it m}$_{t}$ = 0.21, while the
latter each have an anisotropic Fermi contour with cyclotron mass
of ({\it m}$_{l}${\it m}$_{t}$)$^{1/2}$ = 0.47. This is similar to
another system studied previously \cite{vakili04}, except there
the occupied valleys were separated in different parallel layers
while the present system has both within a single 2DES.

We studied two samples (A and B) from the same wafer, each
containing a 50 $\AA$ AlAs QW flanked by Al$_{0.4}$Ga$_{0.6}$As
barriers and modulation doped with Si.  They were patterned in an
L-shaped Hall bar configuration oriented parallel to the in-plane
crystal axes. Ohmic contacts were made by depositing AuGeNi and
alloying in a reducing environment. We also deposited metallic front
and back gates to allow for control of the electron density, {\it
n}. The samples were cooled in a pumped $^3$He refrigerator with a
base temperature of 0.3 K that was equipped with a single-axis
rotating stage to allow for {\it in situ} variation of the angle,
$\theta$, of the magnetic field, {\it B}, with respect to the sample
normal. We define {\it B}$_{\bot}$ as the component of {\it B}
perpendicular to the 2DES plane.

\begin{figure}
\centering
\includegraphics[scale=0.58]{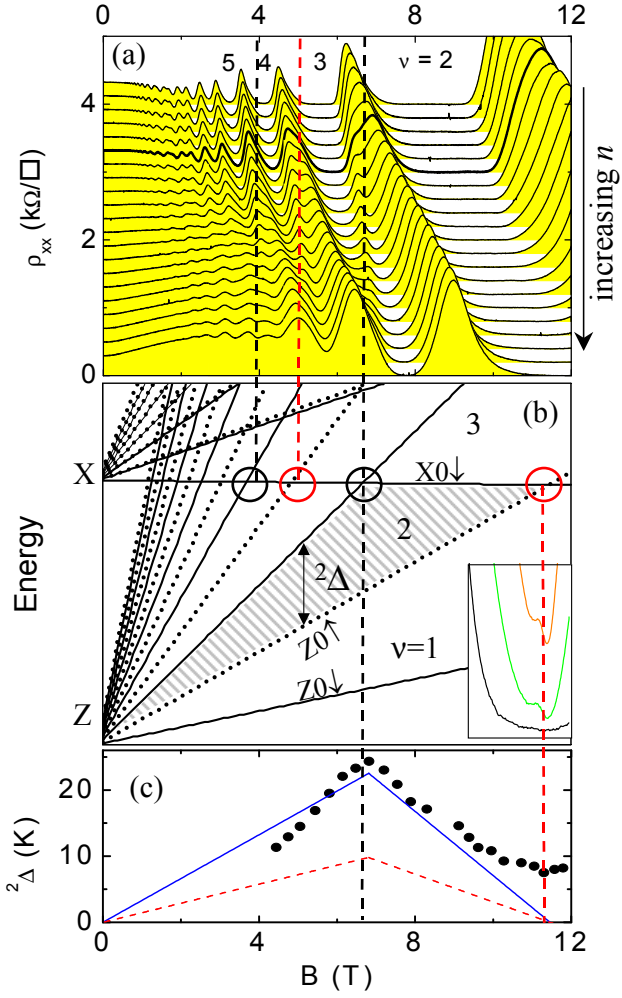}
\caption{(color online) (a) $\rho_{xx}$ vs. {\it B}$_{\bot}$ at
$\theta$ = 0 for {\it n} between 4.0 (top) and 5.6 x 10$^{11}$
cm$^{-2}$ (bottom) in sample A, at {\it T} = 0.3 K. The trace at
the onset of X-valley population ({\it n} = {\it n}$_V$ = 4.2 x
10$^{11}$ cm$^{-2}$) is shown as bold. Traces are vertically
offset for clarity. (b) The energy level "fan" diagram
corresponding to the data shown in (a), with the Z and X valley
LLs labeled and the relevant LL crossings circled (black for
same-spin and red for opposite-spin crossings). The $\nu$ = 2
energy gap ($^2\Delta$) is shaded, and the units of the vertical
axis are arbitrary. The inset of (b) shows $\rho_{xx}$ vs. {\it
B}$_{\bot}$ at {\it n} = 5.5 x 10$^{11}$ cm$^{-2}$ near $\nu$ = 2
[plotted on the same horizontal scale as (a) and (c)] for {\it T}
= 0.3 K (black), 0.95 K (green), and 1.1 K (orange). (c)
$^2\Delta$ vs. {\it B}$_{\bot}$ as {\it n} is varied from 2.1 to
5.7 x 10 $^{11}$ cm$^{-2}$.  The solid blue and dashed red curves
are the predictions for $^2\Delta$ based on the fan diagram of (b)
if all of the enhancement of {\it g}$^*${\it m}$^*$ is assigned to
{\it g}$^*$ or to {\it m}$^*$ respectively.}
\end{figure}

To demonstrate the novel behavior of {\it g}*{\it m}* and {\it g}*
in this system, particularly the equality of the {\it g}* values
when the two valleys are occupied, it is first necessary to
identify the LL crossings and deduce the energy level diagram. In
Fig.\ 1(a), we show longitudinal resistivity ($\rho_{xx}$) vs.
{\it B} at $\theta$ = 0$^o$ in sample A as {\it n} increases from
4.0 (top) to 5.6 x 10$^{11}$ cm$^{-2}$ (bottom). Only the Z valley
is occupied for low values of {\it n}, as evidenced by the
isotropy of $\rho_{xx}$ and by comparison of measured {\it
g}$^*${\it m}$^*$ with values reported in Ref.\ \cite{vakili04b}
where only the Z valley is occupied in a very narrow (45
$\AA$-wide) QW [Fig.\ 2]. At low {\it n}, Shubnikov-de Haas (SdH)
oscillations persist down to {\it B}$_{\bot}$ $\simeq$ 0.5 T. As
{\it n} increases, however, the onset of SdH oscillations moves to
higher {\it B}$_{\bot}$, eventually exceeding 2 T, and $\rho_{xx}$
spikes become evident in the quantum Hall minima (e.g. around
$\nu$ = 3). Both of these observations can be associated with the
population of a second valley, which must be one of the in-plane
valleys (we will refer to it as X) \cite{confine}.  As the X
valley becomes populated, the commingling of its LLs with those of
the Z valley at low {\it B}$_{\bot}$ reduces the size of all
energy gaps and produces the observed shift of SdH onset to higher
{\it B}$_{\bot}$, where the gaps are larger. The spikes, then, can
be associated with crossings that occur between LLs originating
from the X and Z valleys.

This can be shown by considering the energy level (fan) diagram.
The expression for the energy of a LL is:
\begin{equation}
E = (N+\frac{1}{2})\hbar\omega_c \pm \frac{1}{2}g^*\mu_BB \pm
\frac{1}{2}\Delta_V
\end{equation}
where {\it N} is the LL index, $\omega_c$ = {\it e}{\it
B}$_{\bot}$/{\it m}$^*${\it m}$_e$ is the cyclotron frequency, and
$\Delta_V$ is the valley splitting. Since our experiment is only
sensitive to the field positions of the level crossings, which
depend on the relative energy scales in the fan diagram and not on
the overall energy scale, we choose the cyclotron energy of the Z
valley electrons at {\it B}$_{\bot}$ = 1 T as a reference energy.
The expression for the LL energies in these units then becomes:
\begin{equation}
E = (N_i+\frac{1}{2})(m^*_Z/m^*_i)B_{\bot} \pm
\frac{1}{4}g^*_im^*_ZB \pm \frac{h}{4e}n_V
\end{equation}
where {\it n}$_V$ is the threshold density for minority valley
occupation at {\it B}$_{\bot}$ = 0 and {\it i} $\in$ \{X, Z\} is a
valley index.  We deduce {\it n}$_V$ = 4.2 x 10$^{11}$ cm$^{-2}$
from the density where the onset of SdH oscillations begins to
move to higher {\it B}$_{\bot}$ [the bold trace in Fig.\ 1(a)].
This value is also confirmed by the Fourier spectra of the
Shubnikov-de Haas oscillations in our $\rho_{xx}$ traces, which
show additional peaks corresponding to the minority valley above
{\it n}$_V$.  We use the band mass and g-factor ratios, {\it
m}$_Z$/{\it m}$_X$ = 0.45 and {\it g}$_Z$/{\it g}$_X$ = 1, despite
the presence of interactions; we return to this seemingly
questionable assumption later to justify its validity. To complete
the fan diagram, we take {\it g}$^*_Z${\it m}$^*_Z$ = 0.95 to fit
the crossings between opposite spin LLs (same-spin LL crossings
are independent of this, given our assumption of equal g-factors)
\cite{single}. This value of {\it g}$^*_Z${\it m}$^*_Z$ is less
than the measured single-valley value \cite{vakili04b} by
approximately 20$\%$ [Fig.\ 2], but this difference is comparable
to the reduction of {\it g}$^*${\it m}$^*$ reported previously for
a 2DES going from single to double valley occupation
\cite{shkolnikov04}. The resulting fan diagram is exhibited in
Fig.\ 1(b), and we can see that the spikes at $\nu$ = 3, 4, and 5
can reasonably be described as crossings of the Z valley LLs with
the lowest LL of the X valley.

\begin{figure}
\centering
\includegraphics[scale=0.35]{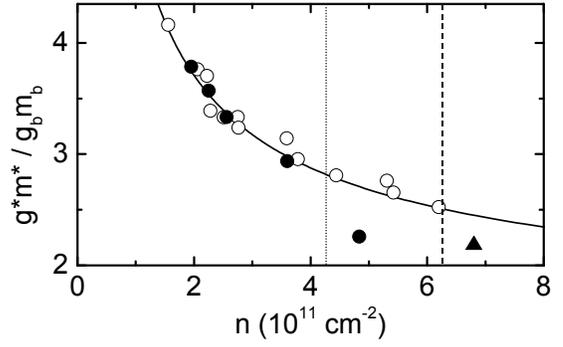}
\caption{The enhancement of {\it g}$^*${\it m}$^*$ over the band
value measured in sample A (filled circles) and sample B (filled
triangle). The open circles are reproduced from Ref.
\cite{vakili04b}, where the electrons occupy only the Z valley in
a very narrow AlAs QW, and the solid line represents the quantum
Monte Carlo prediction of Ref. \cite{attaccalite02}. The vertical
dotted and dashed lines indicate {\it n}$_V$ for samples A and B
respectively.}
\end{figure}

According to Fig.\ 1(b), there should also be a LL crossing
occurring in the $\nu$ = 2 minimum at {\it B}$_{\bot}$ = 11.3 T,
though this is not seen in Fig.\ 1(a).  As the 2DES is heated,
this crossing is indeed revealed near the expected field position
[Fig.\ 1(b) inset]. We have measured the activated energy gap at
$\nu$ = 2, which we call $^2\Delta$ and indicate in Fig.\ 1(b) by
the shaded region,  for temperature, {\it T}, between 0.3 K and 6
K as {\it n} is tuned from 5.7 to 2.2 x 10$^{11}$ cm$^{-2}$. The
results are shown in Fig.\ 1(c).  At the {\it B}$_{\bot}$ = 11.3 T
crossing, $^2\Delta$ is a local minimum, though a finite gap of
7.5 K persists. As {\it n} is reduced, $^2\Delta$ increases until
$\nu$ = 2 reaches the field corresponding to the $\nu$ = 3 LL
crossing ({\it B}$_{\bot}$ = 6.7 T), and decreases thereafter. By
introducing an additional energy scale to the system, namely the
temperature, we can distinguish between the enhancement of {\it
m}$^*$ and {\it g}$^*$ in principle. To generate our fan diagram,
we used {\it g}$^*_Z${\it m}$^*_Z$ = 0.95, which is enhanced by a
factor of about 2.3 over the band value.  In Fig.\ 1(c), we show
the expected gap values if all of this enhancement is assigned to
{\it m}$^*$ (dotted red) or to {\it g}$^*$ (solid blue).  The
latter agrees more closely with the measured gaps and, while we
cannot definitively assign all of the enhancement to {\it g}$^*$,
it seems clear that it is responsible for at least some of the
overall enhancement.  That the agreement persists even below {\it
n}$_V$ ({\it B}$_{\nu=2}$ = 8.7 T) suggests that the
non-equilibrium population of thermally excited minority-valley
electrons continue to affect the overall {\it g}$^*${\it m}$^*$
enhancement. The deviation of the experimental data from the
predicted curve at lower {\it n} likely results from the increase
of {\it g}$^*${\it m}$^*$ that occurs as {\it n} is reduced
\cite{vakili04b, enhance} and as the minority valley is further
depopulated \cite{shkolnikov04}, though the increasing influence
of disorder at low {\it n} may also play a role.

\begin{figure}
\centering
\includegraphics[scale=0.53]{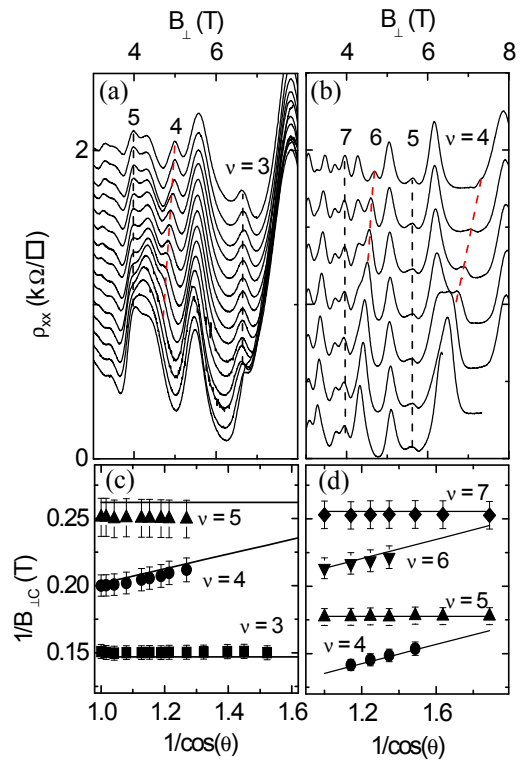}
\caption{(color online) (a) $\rho_{xx}$ vs. {\it B}$_{\bot}$ at n
= 4.9 x 10$^{11}$ cm$^{-2}$ in sample A as the sample is tilted
from $\theta$ = 0 (top) to 52$^o$ (bottom) and (b) at n = 6.8 x
10$^{11}$ cm$^{-2}$ in sample B as the sample is tilted from
$\theta$ = 0 (top) to 58$^o$ (bottom). The positions of the spikes
from (a) and (b) plotted as 1/{\it B}$_{\bot c}$ vs.
1/cos($\theta$) are shown in (c) and (d) respectively. The dotted
lines in (a) and (b) are guides to the eye, and the solid lines in
(c) and (d) are the predicted spike positions based on the 2DES
parameters described in the text. Error bars represent the
full-width at half-maximum of the spikes.}
\end{figure}

In order to justify our assumption that the band mass and g-factor
ratios apply in the presence of interaction, we have tilted our
sample with respect to {\it B}.  Using equation (2), we can write
the expression for the field position of crossings between the Z
valley levels and the lowest LL of the X valley:
\begin{equation}
B_{\bot c} = \frac{\frac{h}{e}n_V}{(2N_Z+1) - \frac{m^*_Z}{m^*_X}
+ g^*_Zm^*_Z(\frac{g^*_X}{g^*_Z} \pm 1)/2cos(\theta)}
\end{equation}
where the plus sign in the $\theta$ dependent term applies for
crossings of LLs with opposite spin and the minus sign for same
spin crossings.  Figure 1(b) indicates that the odd $\nu$
crossings occur between same spin LLs, while the crossings at even
$\nu$ are between opposite spin LLs. In Fig.\ 3(a), we see that
the the $\nu$ = 3 and 5 spike positions remain fixed as we tilt
the sample with respect to {\it B}. The spike positions for $\nu$
= 3, 4, and 5 in sample A are reproduced in Fig.\ 3(c), plotted as
inverse field position vs. 1/cos($\theta$). The lines are the
predicted behavior based on the parameters used to generate the
fan diagram in Fig.\ 1(b), and the flatness of the $\nu$ = 3 and 5
spike positions imply that {\it g}$^*_Z$/{\it g}$^*_X$ = 1 within
an accuracy of better than 1$\%$. We have confirmed this same
behavior in sample B with a larger valley splitting ({\it n}$_V$ =
6.2 x 10$^{11}$ cm$^{-2}$) [Fig.\ 3(b) and (d)], and are again
able to reproduce the spike positions, here using equal values of
{\it g}$^*$ for the two valleys, the band mass ratio, and {\it
g}$^*_Z${\it m}$^*_Z$ = 0.92.  The equality of {\it g}$^*$ for the
two valleys also implies that the band mass ratio applies in our
system, since this ratio determines the positions of the same-spin
crossings in the fan diagram that are consistent with the spike
positions in our data. The fan diagram is not extremely sensitive
to the precise value of the mass ratio, however, so the validity
of using the band value in the presence of interactions cannot be
stated definitively.

Several of our observations regarding the characteristic
parameters of our 2DES are intriguing in light of recent work.
First, it has been shown that {\it g}$^*${\it m}$^*$ is lower in a
degenerate, two-valley 2DES than in the single-valley case
\cite{shkolnikov04}. Our results indicate that {\it g}$^*${\it
m}$^*$ drops significantly with just a modicum of minority valley
occupation [Fig.\ 2]. Second, some measurements performed in Si
metal-oxide-semiconductor field-effect transistors (Si-MOSFETs)
suggest that the enhancement of {\it g}$^*${\it m}$^*$ is
principally due to {\it m}$^*$ enhancement, with {\it g}$^*$ never
more than a factor of 1.5 above the band value \cite{shashkin02}.
In our measurements, however, the factor of 2.3 enhancement of
{\it g}$^*${\it m}$^*$ appears to be attributable, at least in
part, to {\it g}$^*$ enhancement, with {\it m}$^*$ remaining close
to the band value [Fig.\ 1(c)].  We note, however, that in the
range of {\it n} that we have measured $^2\Delta$, reported values
of {\it g}$^*$ and {\it m}$^*$ in Si-MOSFETs are both close to the
band values with the {\it g}$^*$ enhancement even slightly greater
than that of {\it m}$^*$.  It is only at low {\it n} ($\lesssim$ 2
x 10$^{11}$ cm$^{-2}$) that the enhancement of {\it m}$^*$
overtakes that of {\it g}$^*$ in those measurements
\cite{shashkin02}.

Most surprisingly, our data imply that the electrons in each 2DES
are characterized by a {\it single}, enhanced {\it g}$^*$. We
emphasize that this conclusion relies {\it only} on the lack of
angle dependence for the odd $\nu$ spike positions and not on any
other parameters used to fit those positions. The effect of
electron-electron interaction at {\it B}$_{\bot}$ = 0 is often
characterized by the so-called {\it interaction parameter}, {\it
r}$_{s}$, which is equal to the ratio of the Coulomb and Fermi
energies, the latter of which depends on {\it n} and on {\it
m}$^*$ through the density of states \cite{field}. When the X or Z
valleys are singly occupied in QWs with different widths, the
interaction enhancement of the band parameters is different owing
to their different band effective masses and, hence, different
values of {\it r}$_s$ \cite{vakili04b, shkolnikov04}. However,
when these valleys are occupied in the same QW, as in our samples,
the enhancement of {\it g}$^*$ appears to be the same for both
valleys despite having different masses and very different
densities.

Finally, we address the existence of $\rho_{xx}$ spikes at the LL
crossings in our measurements. Such $\rho_{xx}$ spikes have been
reported previously for single-valley 2DESs in tilted magnetic
field at angles where LLs of different spin coincide at the Fermi
energy ({\it E}$_F$) \cite{depoortere00}.  In the case of spin,
LLs are expected to be degenerate in the single-particle case at
all {\it B}$_{\bot}$ for a particular $\theta$. The existence of
persistent gaps and $\rho_{xx}$ spikes are, therefore, quite
surprising and have been explained as a consequence quantum Hall
ferromagnetism \cite{jungwirth98, jungwirth00} and scattering at
boundaries of different spin domains \cite{jungwirth01}.  The
situation is somewhat different in the present case, where the LLs
are only degenerate at particular values of {\it B}$_{\bot}$
[Fig.\ 1(b)].  The variation of the energy gap as {\it E}$_F$
traverses the localized states over a range of {\it B}$_{\bot}$
could yield spike-like features in $\rho_{xx}$ even in the absence
of interaction. On the other hand, it is worth noting that when
the X and Z valley LLs are degenerate, interface roughness
disorder acts as a random, symmetry-breaking field that favors
occupation of the X (Z) valley wherever the QW is locally thicker
(thinner), dividing the 2DES into domains and may contribute
additional dissipation as the electrons scatter at the domain
boundaries.

In summary, we have studied crossings of LLs of a majority Z
valley with the lowest LL of the minority X valley in two AlAs QWs
with different valley splittings. From the positions of the
resistivity spikes and measurements of the energy gap at $\nu$ =
2, we find that {\it g}$^*${\it m}$^*$ drops rapidly upon
occupation of the minority valley.  Of the remaining {\it
g}$^*${\it m}$^*$ enhancement, some or all can be attributed to
{\it g}$^*$ in the {\it n} range of our measurements, and, most
remarkably, that enhancement of {\it g}$^*$ is the same for all
electrons despite the occupation of two very different valleys.

We thank the National Science Foundation for financial support and
E.P.\ De Poortere for discussions.

\break

\end{document}